# Towards Automated Brain Aneurysm Detection in TOF-MRA: Open Data, Weak Labels, and Anatomical Knowledge

Tommaso Di Noto[1] · Guillaume Marie[1] · Sebastien Tourbier[1] · Yasser Alemán-Gómez[1,2] · Oscar Esteban[1] · Guillaume Saliou[1] · Meritxell Bach Cuadra[3] · Patric Hagmann[1] · Jonas Richiardi[1]

## Abstract

Brain aneurysm detection in Time-Of-Flight Magnetic Resonance Angiography (TOF-MRA) has undergone drastic improvements with the advent of Deep Learning (DL). However, performances of supervised DL models heavily rely on the quantity of labeled samples, which are extremely costly to obtain. Here, we present a DL model for aneurysm detection that overcomes the issue with "weak" labels: oversized annotations which are considerably faster to create. Our weak labels resulted to be four times faster to generate than their voxel-wise counterparts. In addition, our model leverages prior anatomical knowledge by focusing only on plausible locations for aneurysm occurrence. We first train and evaluate our model through cross-validation on an in-house TOF-MRA dataset comprising 284 subjects (170 females / 127 healthy controls / 157 patients with 198 aneurysms). On this dataset, our best model achieved a sensitivity of 83%, with False Positive (FP) rate of 0.8 per patient. To assess model generalizability, we then participated in a challenge for aneurysm detection with TOF-MRA data (93 patients, 20 controls, 125 aneurysms). On the public challenge, sensitivity was 68% (FP rate = 2.5), ranking 4th/18 on the open leaderboard. We found no significant difference in sensitivity between aneurysm risk-of-rupture groups ($p = 0.75$), locations ($p = 0.72$), or sizes ($p = 0.15$). Data, code and model weights are released under permissive licenses. We demonstrate that weak labels and anatomical knowledge can alleviate the necessity for prohibitively expensive voxel-wise annotations.

## Introduction

Time-Of-Flight Magnetic Resonance Angiography (TOF-MRA) is a non-invasive and non-contrast imaging technique sensitive to the blood flow in brain vessels. TOF-MRA has found widespread clinical application to identify Unruptured Intracranial Aneurysms (UIAs) which are small (typical diameter ≅ 5 mm) abnormal focal dilatations in cerebral arteries (Chen et al., 2018). If untreated, UIAs can rupture and lead to subarachnoid hemorrhages which have a mortality rate of 40% and usually cause severe disability for patients (Frösen et al., 2012).

Manually assessing a TOF-MRA is a costly process: radiologists detect aneurysms by selectively scrolling through the TOF-MRA volumes in different planes—for instance, they check in the axial plane the most recurrent locations where aneurysms can occur. Then, the sagittal view permits better views of areas like the basilar trunk; afterwards, the coronal view can be used for areas like the anterior cerebral arteries or the Sylvian segments. In addition, Maximum Intensity Projection (MIP) images can be used to search for stenoses, or to confirm potential aneurysms that were spotted.

Considering that the workload of radiologists is steadily increasing (Rao et al., 2021) and the detection of UIAs is a meticulous and non-trivial task (Nakao et al., 2018), the development of automated algorithms that aid clinicians in detecting aneurysms with high sensitivity is an active line of

✉ Tommaso Di Noto
tommaso.di-noto@chuv.ch

[1] Department of Radiology, Lausanne University Hospital and University of Lausanne, Lausanne, Switzerland

[2] Center for Psychiatric Neuroscience, Department of Psychiatry, Lausanne University Hospital and University of Lausanne, Lausanne, Switzerland

[3] Center for Biomedical Imaging, CIBM, Lausanne, Switzerland

Springer

research which holds the promise of improving care while reducing radiologists' assessment times.

Before the popularization of Deep Learning (DL), (Arimura et al., 2004) detected aneurysms by means of image filtering, and later, (Yang et al., 2011) used candidate points of interest in the brain arteries to locate aneurysms. Then, starting from 2016, there was a shift towards DL algorithms, which have now become the de facto standard for UIA detection. Table 1 illustrates several recent studies that use DL for UIA detection. Despite their success, these DL approaches are still constrained by a major bottleneck common to several medical applications: the lack of large, labeled datasets. This is mainly due to two factors: first, the creation of voxel-wise labels for medical images is tedious and time-consuming for radiologists (Razzak et al., 2018); second, none of the TOF-MRA studies to date made their dataset publicly available (Joo et al., 2020; Nakao et al., 2018; Sichtermann et al., 2019; Stember et al., 2019; Ueda et al., 2019). This hampers reproducibility and multi-site analyses that are paramount for building robust DL architectures. The lack of openly available data, such as the TOF-MRA challenge dataset (Timmins et al., 2021), also hinders comparisons across models. Of all reviewed studies of Table 1, only (Baumgartner et al., 2021) evaluated their models on the challenge dataset.

In this work, we develop a fully automated DL network for UIA detection and propose to mitigate the data availability bottleneck as follows: we explore the use of "weak" labels (Abousamra et al., 2020; Ezhov et al., 2018; Ke et al., 2020). These can be coarse or oversized annotations that are less precise, but considerably faster to create for medical experts. In addition, we release our annotated in-house dataset to the community. To the best of our knowledge, this will be the largest openly available TOF-MRA aneurysm dataset to date.

Furthermore, we constrain the DL analysis only to the areas of the brain where aneurysm occurrence is plausible. This anatomically-informed approach aims at simulating the selective analysis that radiologists perform on the TOF-MRA scans. Then, we assess multi-site robustness by evaluating our algorithm on the external TOF-MRA challenge dataset (Timmins et al., 2021). Last, since every aneurysm can have a different prognosis, we investigate how the performances of our model change with respect to aneurysm

**Table 1** Summary of papers that use deep learning models to tackle automated brain aneurysm detection/segmentation

| Paper | Modality | Task(s) | N. Sub | N. Aneurysms | DL Model | Model input | Voxel-wise labels | Use anatomical information | Multi-Site |
|---|---|---|---|---|---|---|---|---|---|
| (Ueda et al., 2019) | MRA | Detection | 1271 | 1477 | ResNet | 2D patches | Not specified | No | Yes |
| (Joo et al., 2020) | MRA | Detection | 744 | 761 | 3D ResNet | 3D patches | Yes | Yes | Yes |
| (Nakao et al., 2018) | MRA | Detection | 450 | 508 | CNN | 2D MIP patches | Yes | Yes | No |
| (Stember et al., 2019) | MRA | Detection | 302 | 336 | RCNN | 2D MIP patches | Yes | No | No |
| (Baumgartner et al., 2021) | MRA | Detection | 254 | N/A | nnDetection | 3D patches | Yes | No | No |
| (Sichtermann et al., 2019) | MRA | Detection (via segmentation) | 85 | 115 | DeepMedic | 3D patches | Yes | Yes | No |
| (Shi et al., 2020) | CTA | Detection + Segmentation | 1177 | 1099 | 3D UNET | 3D patches | Yes | Yes | Yes |
| (Yang et al., 2020) | CTA | Detection | 1068 | 1337 | ResNet | 3D patches | Not specified | No | Yes |
| (Park et al., 2019) | CTA | Segmentation + CAD assessment | 662 | 358 | HeadXNet | 3D patches | Yes | Yes | No |
| (Dai et al., 2020) | CTA | Detection | 311 | 352 | RCNN | 2D NP images | Not specified | No | Yes |
| (Liu et al., 2021) | DSA | Detection + Segmentation | 451 | 485 | 3D UNET | 3D DSA volumes | Yes | Yes | No |
| (Duan et al., 2019) | DSA | Detection | 281 | 261 | 2D CNN | 2D DSA images | Bounding Boxes | Yes | No |
| (Hainc et al., 2020) | DSA | Detection | 240 | 187 | 2D CNN | 2D DSA images | ROI circle | No | No |

Use anatomical information: whether the method uses some sort of anatomical prior knowledge during training, patch sampling or inference (more details in Online Resources – Section A)

*MRA* Magnetic Resonance Angiography, *CTA* Computed Tomography Angiography, *DSA* Digital Subtraction Angiography, *N* number, *Sub* subjects risk-of-rupture groups (defined in "Aneurysm Annotation, Size, Location and Risk Groups for In-house Dataset" section), location and size.

## Materials and Methods

### In-house Dataset

This study was approved by the regional ethics committee; written informed consent was waived. In this retrospective work, we included consecutive patients that underwent TOF-MRA between 2010 and 2015, and for which the corresponding radiological reports were available. Patients with ruptured/treated aneurysms or with other vascular pathologies were excluded. Totally thrombosed aneurysms and infundibula (dilatations of the origin of an artery) were likewise excluded. In total, we retrieved 284 TOF-MRA subjects: 157 had one (or more) UIAs, while 127 did not present any. Table 2 illustrates the main demographic information for our study group. A 3D gradient recalled echo sequence with Partial Fourier technique was used for all subjects (acquisition parameters are reported in Online Resources—Table 1). 214 subjects of this study were also used in (Di Noto et al., 2020). This prior article dealt with patch-wise classification, whereas here we address patient-wise aneurysm detection. The dataset was anonymized and organized according to the Brain Imaging Data Structure (BIDS) standard (Gorgolewski, 2008). It is available on OpenNeuro (Markiewicz et al., 2021) at https://openneuro.org/datasets/ds003949 under the CC0 license.

### Aneurysm Annotation, Size, Location and Risk Groups for In-house Dataset

Aneurysms were annotated by one radiologist with 2 years of experience in neuroimaging, and double-checked by a senior neuroradiologist with over 15 years of experience to exclude potential false positives or false negatives. Two annotation schemes were followed:

1. **Weak labels**: for most subjects (246/284), the radiologist used the Multi-image Analysis GUI (Mango) software (v. 4.0.1) to create the aforementioned weak labels. These correspond to spheres that enclose the whole aneurysm, regardless of the shape. In other words, the size of the spheres was chosen manually by our radiologist on a case-by-case basis ensuring that the whole aneurysm was always entirely enclosed within the sphere. A visual example of one weak label is provided in Fig. 1.
2. **Voxel-wise labels**: for the remaining subjects (38/284), the radiologist used ITK-SNAP (v. 3.6.0) (Yushkevich et al., 2006) to create voxel-wise labels drawn slice by slice scrolling in the axial plane. No specific selection criterion was used to select the 38 subjects, which were consecutive to the 246 of the first group.

The overall number of aneurysms included in the study is 198 (178 saccular, 20 fusiform). Table 3 shows their locations and sizes grouped according to the PHASES score (Greving et al., 2014). This is a clinical score used to assess the 5-year risk of rupture of aneurysms. Although using the PHASES sizes leads to a very skewed distribution (e.g. the category size $d \leq 7$ mm contains 91% of the aneurysms), we decided to stick to this grouping since it is the one used in the clinic.

In addition, for post-hoc analysis and stratification purposes, we divided the aneurysms into two groups based on their risk of rupture: *low-risk* and *medium-risk*. Aneurysms in the *low-risk* group are those that are monitored over time, but do not require any intervention. Instead, aneurysms in the *medium-risk* group can be considered for treatment. We computed for each aneurysm a partial PHASES score that only considered size, location, and patient's age, thus neglecting population, hypertension, and earlier aneurysmal hemorrhage, since this information was not available for all patients. If an aneurysm had partial PHASES score $\leq 4$, it was assigned to the *low-risk* group, while if it had a partial score $> 4$, it was assigned to the *medium-risk* group. Each aneurysm was reviewed by our senior neuroradiologist to assess whether the partial PHASES score was reasonable.

**Table 2** Demographics of the study sample

|  | Patients | Controls | Test, $p$ value | Whole Sample |
|---|---|---|---|---|
| **N** | 157 | 127 | / | 284 |
| **Age (y)** | $56 \pm 14$ | $46 \pm 17$ | $t = -4, 3, p = 7.6 \times 10^{-7}$ | $51 \pm 16$ |
| **Sex** | 53 M, 104F | 61 M, 66F | $\chi^2 = 5.9\ p = 0.01$ | 114 M, 170F |
| **# UIA** | 198 | 0 | / | 198 |

Patients = subjects with aneurysm(s). Controls = subjects without aneurysms. Age calculated in years and presented as mean ± standard deviation. Two-sided t-test to compare age between patients and controls. Chi-squared test to compare sex counts between patients and controls

*N* number of samples, *M* males, *F* females, *UIA* Unruptured Intracranial Aneurysms

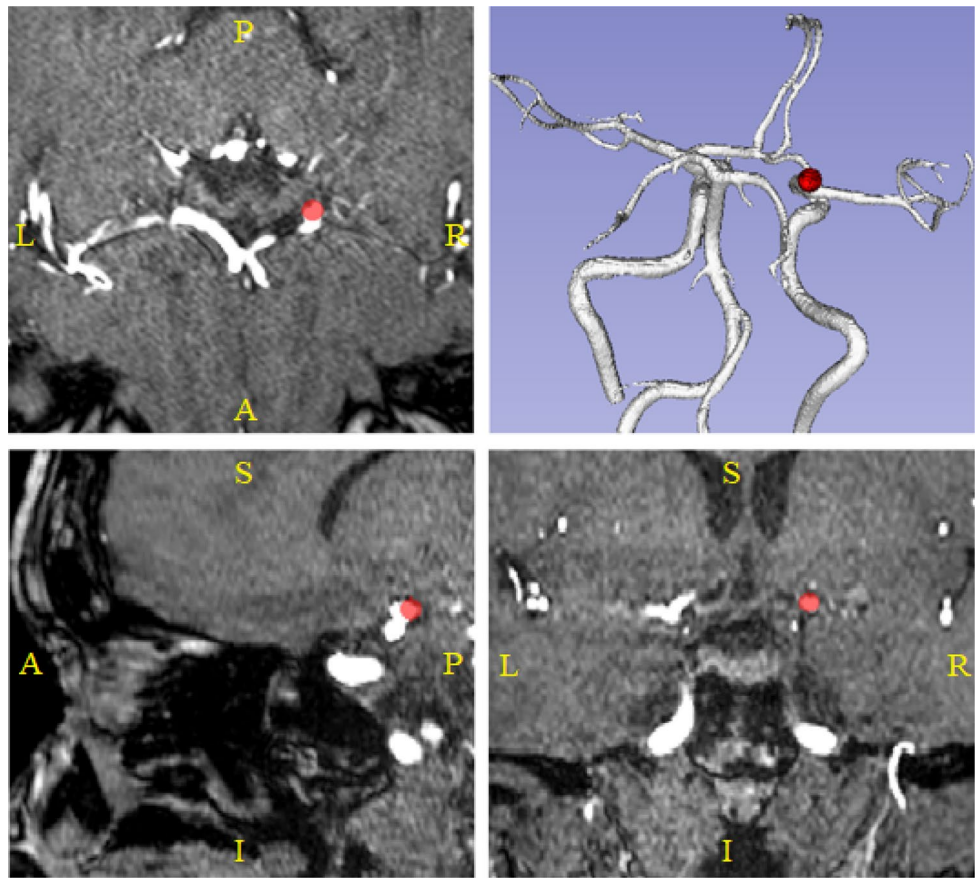

**Fig. 1** TOF-MRA orthogonal views of a 62-year-old female patient. Red areas correspond to our spherical weak labels. Top-left: axial plane; top-right: 3D posterior reconstruction of the cerebral arteries; bottom-left: sagittal plane; bottom-right: coronal plane

Fusiform aneurysms were excluded from the risk analysis since the PHASES score was built for saccular aneurysms. Similarly, extracranial carotid artery aneurysms were excluded since they do not bleed in the subarachnoid space. This resulted in 141 *low-risk* and 23 *medium-risk* aneurysms. A table summarizing aneurysm shape, size, location, associated PHASES score and risk groups is provided as Supplementary Material.

**Table 3** Locations and sizes of aneurysms according to the PHASES score for the in-house dataset

|  |  | Count | % |
|---|---|---|---|
| **Location** | ICA | 59 | 29.8 (59/198) |
|  | MCA | 57 | 28.8 (57/198) |
|  | ACA/Pcom/Posterior | 82 | 41.4 (82/198) |
| **Size** | $d \leq 7$ mm | 180 | 91.0 (180/198) |
|  | 7 – 9, 9 mm | 7 | 3.5 (7/198) |
|  | 10 – 19, 9 mm | 10 | 5.0 (10/198) |
|  | $d \geq 20$ mm | 1 | 0.5 (1/198) |

*ICA* Internal Carotid Artery, *MCA* Middle Cerebral Artery, *ACA* Anterior Cerebral Arteries, *Pcom* Posterior communicating artery, *Posterior* posterior circulation, *d* maximum diameter

### Data Processing

Several preprocessing steps were carried out for each subject. First, we performed skull-stripping with the FSL Brain Extraction Tool (v. 6.0.1) (Smith, 2002). Second, we applied N4 bias field correction with SimpleITK (v. 1.2.0) (Tustison et al., 2010). Third, we resampled all volumes to a median voxel spacing, again with SimpleITK. This effectively normalizes nonuniform voxel sizes (Isensee et al., 2021). Last, a probabilistic vessel atlas built from multi-center MRA datasets (Mouches & Forkert, 2014) was co-registered to each patient's TOF-MRA using ANTS (v. 2.3.1) (Avants et al., 2014) (details in Online Resources – Section B). The atlas was used both during training and inference (see "Use of Anatomical Information" section).

### Network, Cross-Validation, Metrics and Statistics

**Network** The deep learning model used in this study is a custom 3D UNET, inspired by the original work (Özgün et al., 2016). We used upsample layers in the decoding branch rather than transpose convolutions since these led to faster model convergence. Figure 2 illustrates the structure of our network. We used 3D TOF-MRA patches as input to our network. We set the side of the input patches to 64x64x64 voxels to include even the largest aneurysms. All patches were Z-score normalized, as is common practice (Bengio et al., 2016). A kernel size of 3x3x3 was used in all convolutional layers, with padding and stride = 1. We applied the ReLU activation function for all layers, except for the last layer which is followed by a sigmoid function. To fit the model, the Adam optimization algorithm (Kingma & Ba, 2015) was applied with adaptive learning rate (initial learning rate = 0.0001). We trained the model for 100 epochs, and we adopted the Combo loss function (Taghanaki et al., 2019) with $\alpha = \beta = 0.5$. This function combines Dice and Cross-entropy, and has proven to be effective for imbalanced segmentation tasks. We used Xavier initialization (Glorot & Bengio, 2010) for all layers. Biases were initialized to 0 and a batch size of 8 was chosen. Batch normalization (Ioffe & Szegedy, 2015) was used to prevent overfitting. The number of convolutional filters, the batch size, the value of $\alpha$ (and therefore $\beta = 1 - \alpha$) and the learning rate were chosen using the Optuna algorithm (Akiba et al., 2019) on an internal validation set (20% of training cases of external cross-validation fold 1, see below for cross-validation details). The total number of trainable parameters in our network is 855,111. Training and evaluation were performed with Tensorflow 2.4.0 and a GeForce RTX 2080TI GPU with 11 GB of SDRAM.

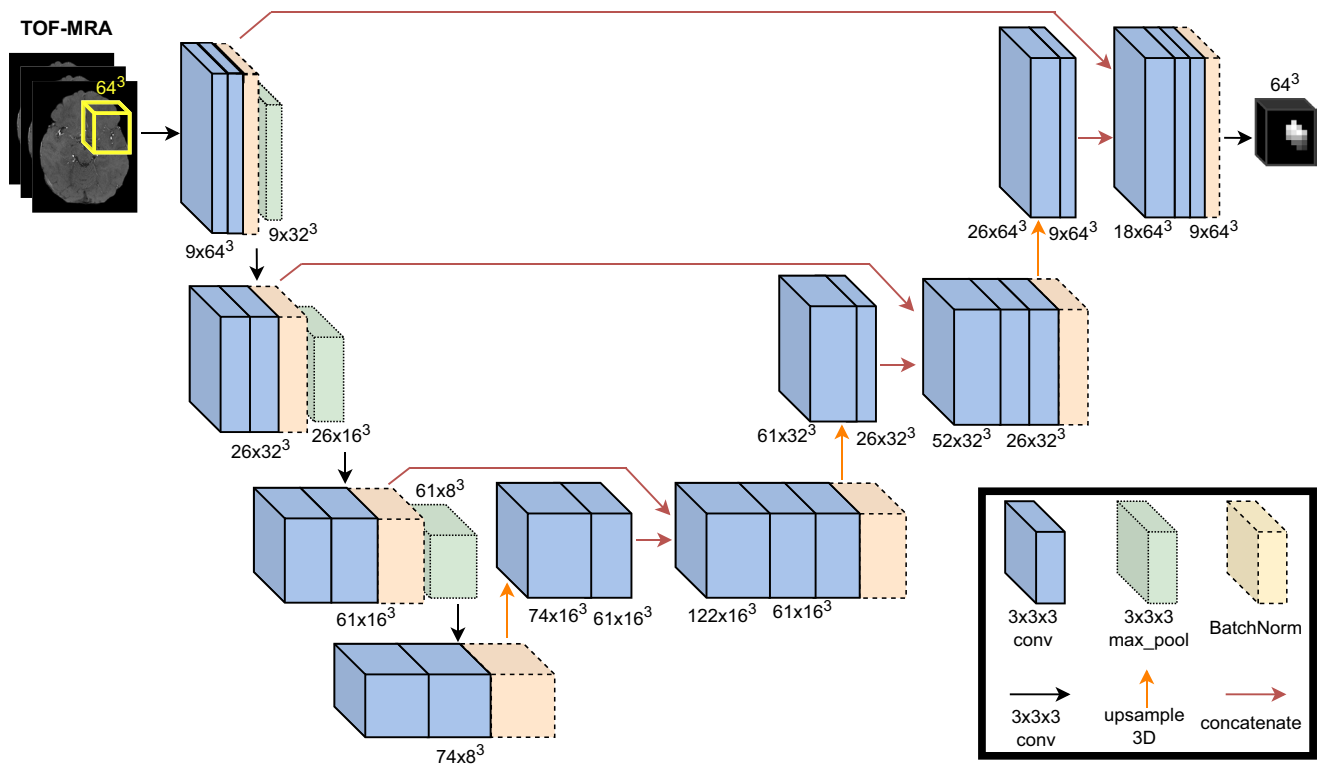

**Fig. 2** Proposed variant of the 3D UNET. The input corresponds to a 64x64x64 voxels TOF-MRA patch. The output is a probabilistic patch with the same size of the input, but where each voxel corresponds to the probability of either belonging to foreground (i.e., aneurysm) or background. *Conv* convolutional, *Max_pool* max pooling, *BatchNorm* batch normalization

**Cross-validation** To evaluate detection performances, we conducted a fivefold cross-validation on the 246 subjects with weak labels. At each cross-validation split, 80% ($\approx$197/246) of the subjects are used for training the network, while the remaining 20% ($\approx$49/246) of the subjects are used to compute patient-wise results (i.e. for inference). This division occurs 5 times (as the number of folds) and every time a different 80%-20% split is created, meaning that all 246 patients are ultimately used for evaluation. At each cross-validation split, the 38 patients with voxel-wise labels were always added to the training set to increase the effect size of label quality in further analyses (see experiments in "Use of Weak Labels"). To avoid over-optimistic results, we ensured that patients with multiple sessions were not split between training and test set. In order to make results comparable across experiments, we always used the same cross-validation split and we released this split for reproducibility on https://github.com/connectomicslab/Aneurysm_Detection.

In all experiments on the in-house dataset, we always pre-trained our network on the whole ADAM training dataset (Timmins et al., 2021) and then fine-tuned it on the in-house training data. To validate the effectiveness of pre-training on ADAM, we performed ablation experiments of domain adaptation across the two datasets (in-house and ADAM). As these experiments are beyond the main focus of the manuscript, we added them in the Online Resources – Section F. When performing pre-training on the ADAM dataset, we applied both anatomically-informed expedients described below in "Use of Anatomical Information" section.

**Metrics and Statistics** In line with the ADAM challenge (presented in " Participation to the ADAM Challenge" section), we used sensitivity and false positive (FP) rate as detection metrics. A detection was considered correct if the center-of-mass of the predicted aneurysm was located within the maximum aneurysm size of the ground truth mask. In addition, we computed the Free-response Receiver Operating Characteristic (FROC) curve (Chakraborty & Berbaum, 2004). To compare different model configurations, we used a two-sided Wilcoxon signed-rank test of the areas under the FROC curves across test subjects, as similarly performed in (Ward et al., 1999). To compare the performances of a configuration with respect to aneurysm rupture risk, location and size we performed several Chi-squared tests (McHugh, 2012). The statistical tests were performed using SciPy (v.1.4.1), setting a significance threshold $\alpha = 0.05$.

## Experiments

In this section, we will present the four experiments that we conducted: in "Use of Weak Labels" section, we investigate the use of weak labels in terms of difference in annotation time and in detection performances, when comparing to voxel-wise labels; in "Use of Anatomical Information" section, we present our anatomically-informed approach for tackling UIA detection; in "Participation to the ADAM Challenge" section, we describe the participation to the ADAM challenge to investigate the generalization of our model; in "Performances With Respect to Risk-of-rupture, Location and Size" section, we analyze the changes in detection performances with respect to aneurysm risk-of-rupture groups, location and size.

### Use of Weak Labels

The goal of this experiment was to answer the following questions: 1) how much faster is the creation of weak labels with respect to the creation of voxel-wise labels? 2) what is the impact of using weak labels in terms of detection performances when comparing to voxel-wise labels?

To answer the first question, we selected a subset of 14 patients (mean aneurysm size (s.d.) = 5.2 (1.0) mm), and we assessed the time difference between the two annotation schemes (i.e. all 14 patients were annotated first with weak labels, and then with voxel-wise labels). These 14 patients were chosen randomly among the 284 TOF-MRA subjects, but we ensured that the mean aneurysm size was representative of the whole cohort.

To answer the second question, we used the 38 subjects with voxel-wise labels and for these patients we artificially generated corresponding weak spherical labels ('weakened' labels, details in Online Resources – Section C). Then, to evaluate the influence of annotation quality (weakened vs. voxel-wise) in terms of detection performances, we conducted 3 experiments in which we used an increasing number of patients with voxel-wise labels: (i) all 38 patients with weakened labels (Model 1, Table 4), (ii) 19 patients with weakened labels and 19 with voxel-wise labels (Model 2, Table 4), and (iii) all 38 patients with voxel-wise labels (Model 3, Table 4). Results related to the use of weak labels are presented in "Weak Labels Allow Fourfold Annotation Speedup Without Degrading Performances" section.

### Use of Anatomical Information

Because the task of aneurysm detection is extremely spatially constrained, we exploit the prior information that aneurysms a) must occur in vessels, and b) tend to occur in specific locations of the vasculature. To include this anatomical knowledge, one of our radiologists pinpointed in the vessel atlas (described in "Aneurysm Annotation, Size,

**Table 4** Average detection results on the in-house dataset across test folds when changing the ratio of voxel-wise/*weakened* labels. Sensitivity values are reported as mean and 95% Wilson confidence interval inside parentheses

| Model Configu-ration | Anatomically-informed patch selection | Anatomically-informed sliding window | Labels of 38 added subs | Avg. Sensitivity (CI) | Avg. FP rate |
|---|---|---|---|---|---|
| *Model 1* | Yes | Yes | 38 *weakened* | 95/127 = 75% (65%, 80%) | 1.3 |
| *Model 2* | Yes | Yes | 19 *weakened*, 19 voxel-wise | 99/127 = 78% (68%, 82%) | **0.9** |
| *Model 3* | Yes | Yes | 38 voxel-wise | 101/127 = **80%** (72%, 85%) | 1.2 |

Bold values represent the best performances

*Avg* average, *FP* false positive, *CI* confidence interval, *voxel-wise* labels drawn slice by slice on the axial plane, *weakened* voxel-wise labels that are artificially converted to weak spherical labels, *subs* subjects

Location and Risk Groups for In-house Dataset" section) the location of 20 landmark points where aneurysm occurrence is most frequent (list in Online Resources – Table 2). These points were chosen according to the literature (Brown & Broderick, 2014) and were co-registered to the TOF-MRA space of each subject, as illustrated in Fig. 3.

**Training** We apply an anatomically-informed selection of training patches to sample both negative (without aneurysms) and positive (with aneurysms) samples. Specifically, 8 positive patches per aneurysm were randomly extracted in a non-centered fashion. Then, we extracted 50 negative patches per TOF-MRA volume. Out of these, 20 were centered in correspondence with the landmark points, 20 were patches containing vessels (details in Online Resources – Section D), and 10 were extracted randomly. Overall, this sampling strategy allows us to extract most of the negative patches (i.e., all but the random ones) which are comparable to the positive ones in terms of average intensity. To mitigate class imbalance, we applied data augmentations on positive patches: namely, rotations (90°, 180°, 270°), flipping (horizontal, vertical), contrast adjustment, gamma correction, and addition of gaussian noise.

**Inference** The patient-wise evaluation was performed following the sliding window approach (details in Online Resources – Section E). We exploited again the prior anatomical information described above by retaining only the patches which are both within a minimum distance from the landmark points and fulfill specific intensity criteria (details in Online Resources – Section D). The rationale behind this choice was to only focus on patches located in the main cerebral arteries, as shown in Fig. 4. Two post-processing steps were adopted: first, we kept a maximum of 5 candidate aneurysms per patient (only the 5 most probable). Second, we applied test-time augmentation to increase sensitivity.

**Validation** To validate the effectiveness of our two anatomically-informed expedients (patch sampling and sliding window), we first evaluated an anatomically-agnostic baseline where none of the two expedients is used and the 38 added subjects have *weakened* labels (*Model 4*, Table 5). Second, we evaluated the same anatomically-agnostic baseline (none of the two expedients used) but with the 38 subjects having voxel-wise labels (*Model 5*, Table 5). Third, we tested one model where only the anatomically-informed patch sampling is carried out (*Model 6*, Table 5). Last, we computed performances when only the anatomically-informed sliding window is performed (*Model 7*, Table 5). Results related to the use of anatomical information are shown in "Anatomically-informed Sliding Window Increases Detection Performances" section.

**Participation to the ADAM Challenge**

To evaluate model performances in data coming from a different institution, we participated to the Aneurysm Detection And segMentation (ADAM) challenge (http://adam.isi.uu.nl/) (Timmins et al., 2021). The ADAM training dataset is composed of 113 TOF-MRA (93 patients with UIAs, 20 controls). The total number of UIAs is 125 and the voxel-wise annotations were drawn in the axial plane by two radiologists. Instead, the unreleased test dataset is made of 141 cases (117 patients, 26 controls) and it is solely used by the organizers to compute patient-wise results. To improve detection performances on the ADAM test set, we pre-trained our network on the whole in-house dataset and then fine-tuned it on the ADAM training dataset. Results related to our model submission to the ADAM challenge are presented in "The Proposed Model Ranked At the Top of the ADAM Challenge" section.

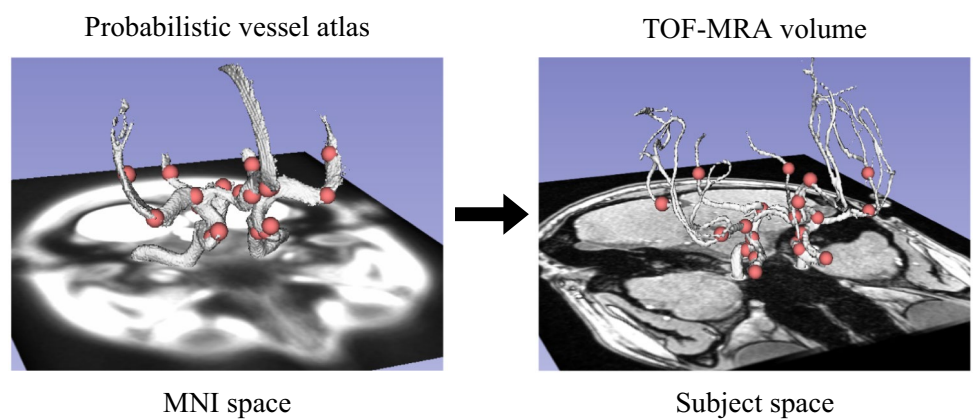

**Fig. 3** **left**: 20 landmark points (in red) located in specific positions of the cerebral arteries (white segmentation) in MNI space. **right**: same landmark points co-registered to the TOF-MRA space of a 21-year-old, female subject without aneurysms

Probabilistic vessel atlas     TOF-MRA volume

MNI space     Subject space

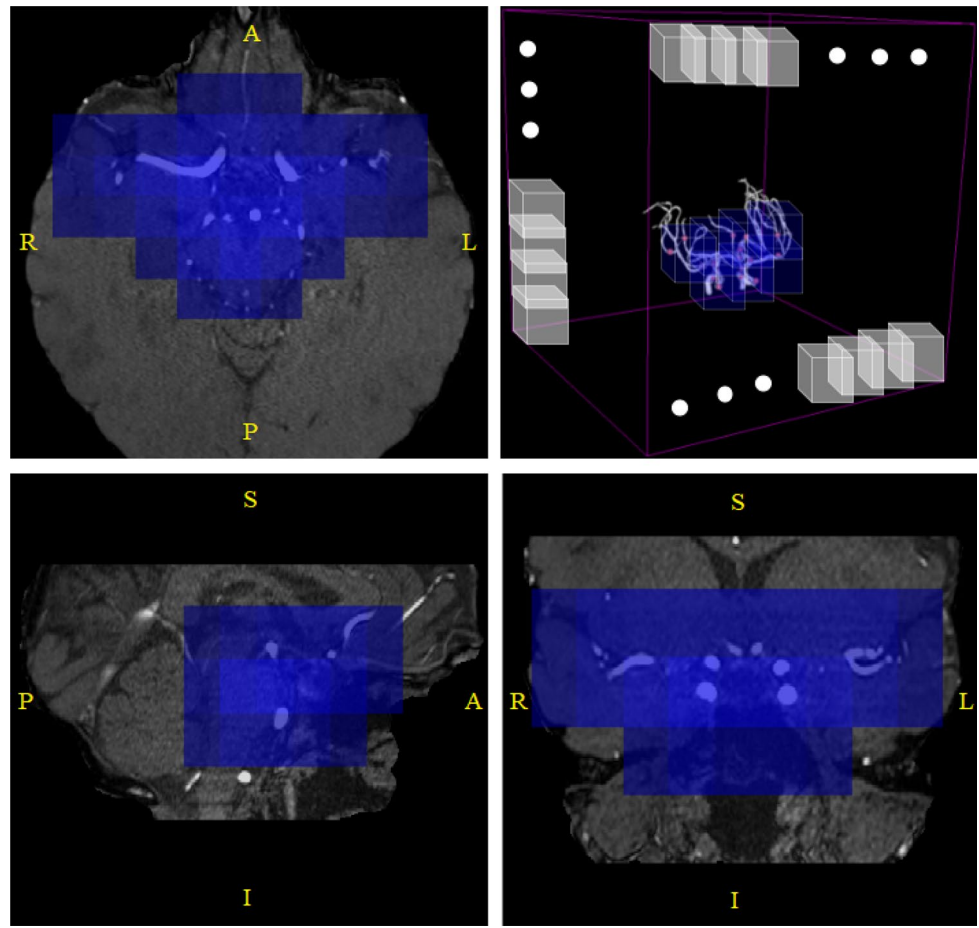

**Fig. 4** TOF-MRA orthogonal views of a 62-year-old female subject after brain extraction: blue patches are the ones which are retained in the *anatomically-informed* sliding-window approach. (top-right): 3D schematic representation of sliding-window approach; out of all the patches in the volume (white patches), we only retain those located in the proximity of the main brain arteries (blue ones)

### Performances with Respect to Risk-of-rupture, Location and Size

Each aneurysm has a different prognosis and, depending on its risk-of-rupture group (defined in "Aneurysm Annotation, Size, Location and Risk Groups for In-house Dataset" section), it will be either monitored over time (low risk) or considered for treatment (medium risk). Therefore, we investigated how detection performances would vary with respect to the risk-of-rupture groups. In addition, we also explored how performances would vary with respect to aneurysm location and size. Although the latter analysis is less relevant from a clinical perspective, it is still interesting from a methodological point of view and it is also frequent in the literature. Results related to the detection performances with respect to aneurysm risk-of-rupture groups, location and size are described in "Detection Performances Across Rupture Risk, Location, and Size" section.

**Table 5** Average detection results on the in-house dataset across test folds when applying none, or one of the two anatomically-informed expedients. Sensitivity values are reported as mean and 95% Wilson confidence interval inside parentheses

| Model Configuration | Anatomically-informed patch selection | Anatomically-informed sliding window | Labels of 38 added subs | Avg. Sensitivity (CI) | Avg. FP rate |
|---|---|---|---|---|---|
| *Model 4* | No | No | 38 *weakened* | 83/127 = 65% (55%, 71%) | 4.6 |
| *Model 5* | No | No | 38 voxel-wise | 95/127 = 74% (63%, 78%) | 4.5 |
| *Model 6* | Yes | No | 38 voxel-wise | 61/127 = 48% (38%, 55%) | 4.8 |
| *Model 7* | No | Yes | 38 voxel-wise | 106/127 = **83%** (75%, 88%) | **0.8** |

Bold values represent the best performances

*Avg* average, *FP* false positive, *CI* confidence interval, *voxel-wise* labels drawn slice by slice on the axial plane, *weakened* voxel-wise labels that are artificially converted to weak spherical labels, *subs* subjects

# Results

## Weak Labels Allow Fourfold Annotation Speedup Without Degrading Performances

When measuring the time needed to create weak vs. voxel-wise annotations on the 14 subjects described in "Use of Weak Labels" section, we noticed a significant difference (two-sided Wilcoxon signed-rank test – annotation timings, $W = 0$, $p = 0.001$): creating weak annotations (average 23 s $\pm$ 6 per aneurysm) resulted to be approximately 4 times faster than creating voxel-wise annotations (average 93 s $\pm$ 25). A more detailed stratification of the timings with respect to location and size is provided in Supplementary Figs. 1 and 2.

Subsequently, to investigate the effect that voxel-wise labels can have for detection performances with respect to weak labels, we conducted several experiments where an increasing ratio of voxel-wise/*weakened* labels was used for the 38 patients described in "Use of Weak Labels" section. Table 4 shows detection performances when the ratio is gradually increased.

The configuration with all voxel-wise labels (*Model 3*) had higher sensitivity with respect to the other two configurations with *weakened* labels (*Model 1* and *Model 2*). However, this difference was not significant (two-sided Wilcoxon signed-rank test on the areas under the FROC curves, $W = 14.0$, $p = 0.054$ when comparing to *Model 1* and $W = 685.5$, $p = 0.977$ when comparing to *Model 2*).

## Anatomically-informed Sliding Window Increases Detection Performances

In Table 5, we report detection results when adopting zero, one, or both anatomically-informed expedients presented in "Use of Anatomical Information" section. In the anatomically-agnostic baseline with the 38 subjects having *weakened* labels (*Model 4*), the negative patch sampling is random and all non-zero patches of the TOF-MRA volumes are retained in the sliding window approach, thus disregarding any anatomical constrain. Similarly, row 2 (*Model 5*) shows detection results when using neither the anatomically-informed patch sampling nor the anatomically-informed sliding window, but this time with the 38 subjects having voxel-wise labels. Row 3 (*Model 6*) illustrates detection performances when only the anatomically-informed patch sampling is applied, but the sliding window is still anatomically-agnostic. Instead, row 4 (*Model 7*) shows the inverse scenario (i.e. random negative patch sampling, but anatomically-informed sliding window). *Model 7* statistically outperformed *Model 5* ($W = 74.5$, $p = 2 \times 10^{-6}$), thus indicating that the anatomically-informed sliding window is an effective expedient to increase detection results. In fact, sensitivity is increased and the average FP rate is drastically reduced. In addition, we compared *Model 5* and *Model 6* and we saw that *Model 5* significantly outperforms *Model 6* ($W = 202.0$, $p = 8 \times 10^{-6}$). This finding shows that the anatomically-informed patch sampling is detrimental for detection performances when the sliding window is anatomically-agnostic. Last, when comparing *Model 3* and *Model 7* we found no significant difference ($W = 81.5$, $p = 0.24$): this result indicates that the anatomically-informed patch sampling is not detrimental when we are also applying the anatomically-informed sliding window.

To provide a visual interpretation of our network predictions, we show in Fig. 5 one correctly identified aneurysm (true positive), one small, missed aneurysm (false negative) and one false positive prediction. Also, in Fig. 6 we report the FROC curves corresponding to *Model 5*, *Model 6*, and *Model 7*. This figure reflects the statistical tests: *Model 7* (green curve) outperforms the anatomically-agnostic *Model 5* (red curve) at all operating points. Similarly, *Model 5* (red curve) significantly outperforms *Model 6* (blue curve), confirming the effectiveness of the anatomically-informed sliding window and the ineffectiveness of the anatomically-informed patch sampling.

## The Proposed Model Ranked At the Top of the ADAM Challenge

Table 6 illustrates detection results on the ADAM test dataset. Our algorithm ranked in 4th/18 position for detection and in 4th/15 position for segmentation (with highest volumetric similarity). Interested readers can check the methods proposed by other teams on the challenge website (https://adam.isi.uu.nl/) and in the paper (Timmins et al., 2021).

## Detection Performances Across Rupture Risk, Location, and Size

Supplementary Fig. 3 illustrates performances achieved by one of our top-performing models (*Model 3*, Table 4) stratified according to the two risk groups presented in "Aneurysm Annotation, Size, Location and Risk Groups for In-house Dataset" section. For the *low-risk* group, our model reaches a mean sensitivity of 80%, while for the *medium-risk* group it reaches a mean sensitivity of 73%. The difference was not significant ($\chi^2 = 0.09$, $DoF = 1$, $p = 0.75$). In Supplementary Figs. 4 and 5, we also report the model sensitivity stratified according to aneurysm location and size of the PHASES score, respectively. No significant difference was found across different locations ($\chi^2 = 0.64$, $DoF = 2$, $p = 0.72$) or sizes ($\chi^2 = 0.92$, $DoF = 2$, $p = 0.15$, excluding $n = 1$ huge aneurysm with $_s > 20$ mm). Regarding aneurysm

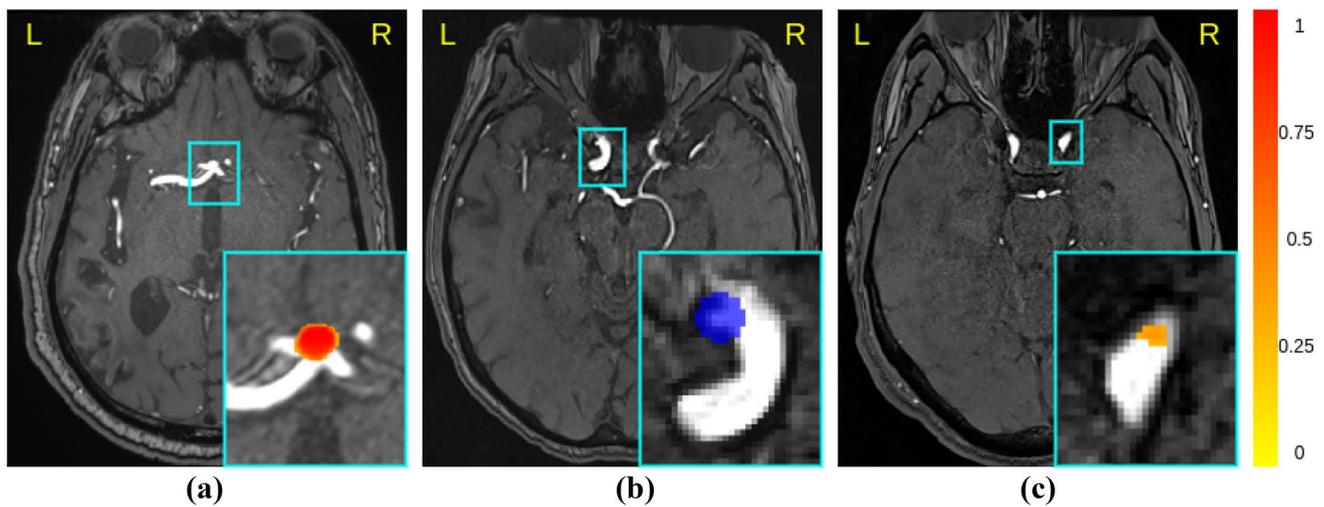

**Fig. 5** Qualitative analysis of predictions and errors. The heatmap generated by the network ranges from 0 (low probability, yellow color) to 1 (high probability, red color) (**a**) True positive prediction in the anterior communicating artery. **b** False negative (i.e., missed aneurysm) in the internal carotid artery. The ground truth label mask is shown in blue. **c** False positive prediction in the internal carotid artery

size, we conducted a further stratification of performances since most of the aneurysms lied in the group (< 7 mm). Thus, we divided this group into subgroups, namely ≤ 3, 3 < s ≤ 5, and 5 < s < 7. Detection results with this more granular stratification are shown in Supplementary Fig. 6. The model sensitivity was significantly lower for the tiny aneurysms (≤ 3) with respect to the other two subgroups ($\chi^2 = 27.57$, $DoF = 2$, $p = 10^{-6}$).

## Discussion

This work shows that competitive results can be obtained in automated aneurysm detection for TOF-MRA data even with rapid data annotation. To this end, we proposed a fully-automated, deep learning algorithm that is trained using weak labels and exploits prior anatomical knowledge.

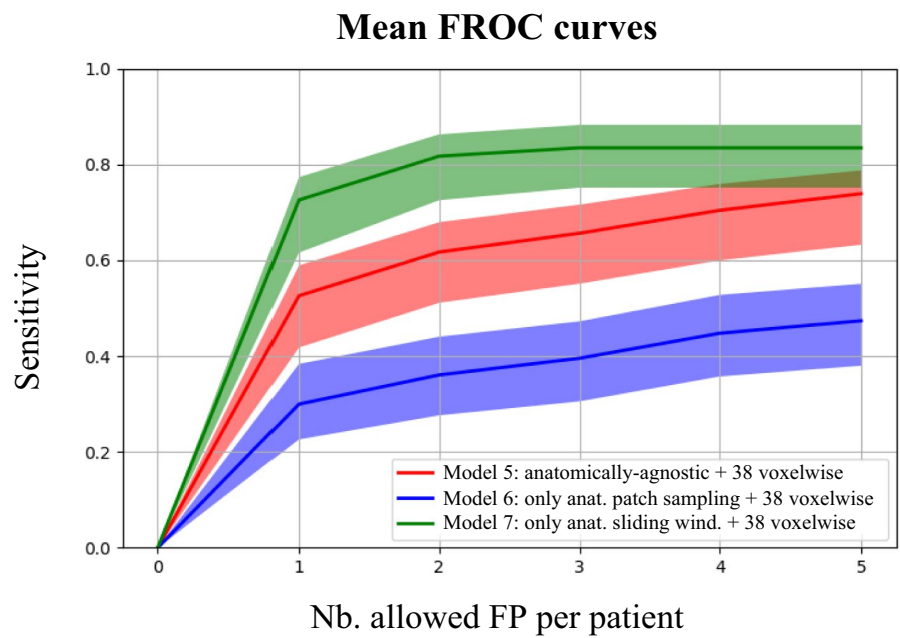

**Fig. 6** Mean Free-response Receiver Operating Characteristic (FROC) curves across the five test folds of the cross-validation. Shaded areas represent the 95% Wilson confidence interval. The three models correspond to *Model 5*, *Model 6*, and *Model 7*. *Anatomically-agnostic model:* none of the two anatomically-informed expedients are used. *Anat:* Anatomically-Informed

**Table 6** Detection results on the ADAM dataset. Our team (in bold) ranked in 4th position in the open leaderboard out of 18 participating groups

| Ranking | Team | Detection | |
|---|---|---|---|
| | | Sens | Avg. FP rate |
| 1 | abc | 68% | 0.40 |
| 2 | xlim | 70% | 4.03 |
| 3 | mibaumgartner | 67% | 0.13 |
| **4** | **unil-chuv3** | **68%** | **2.50** |
| 5 | joker | 63% | 0.16 |
| … | | | |
| 18 | ibbm | 2% | 0.01 |

*Sens* sensitivity, *FP* false positive

Despite being less accurate, weak labels are drastically faster to create for medical experts reducing fourfold the annotation time. Although the configuration with all voxel-wise labels (*Model 3*, Table 4) had higher sensitivity, we found no statistical difference when comparing with the configurations with some (*Model 2*) or all *weakened* labels (*Model 1*). This finding indicates that weak labels are sufficient to obtain satisfactory detection results on our in-house dataset. If reasoning in terms of larger datasets (e.g., thousands of patients), the weak annotation proposed in this work is a scalable solution which can significantly alleviate the annotation bottleneck in medical ML applications.

In addition to the use of weak labels, our model leverages the underlying anatomy of the brain vasculature (i.e., we "*anatomically-informed*" our network) in order to simulate the radiologists' exploration of the TOF-MRA scans. First, most of the negative patches (i.e. patches without aneurysms) extracted during training either contained a vessel or were located in correspondence with the aneurysm landmark points. Second, we limited the sliding window approach only to regions of the brain that are plausible for aneurysm occurrence. These constraints reflect the radiologists' behavior in the sense that only regions containing vessels, or at higher risk for aneurysms are scanned, while the rest of the brain is neglected. The experiments in "Anatomically-informed Sliding Window Increases Detection Performances" section showed that the anatomically-informed sliding window is an effective expedient since it increases sensitivity, while reducing the average FP rate. Instead, the anatomically-informed patch sampling proved to be negligible when combined with the anatomically-informed sliding-window (*Model 3 vs. Model 7*), or even detrimental when the sliding window was anatomically-agnostic (*Model 5 vs. Model 6*). We hypothesize that applying only the anatomically-informed patch sampling leads to a domain shift issue: specifically, the model is trained using intensity-matched patches, but then is tested with any patch in the brain (because there is no anatomically-informed sliding window). We think this difference between training and test domain is what causes the decrease in performances in the comparison *Model 5* vs. *Model 6*.

Nevertheless, the anatomically-informed sliding window expedient suggests that injecting prior anatomical knowledge in the pipeline can improve detection performances. We believe this general principle is also applicable to other pathologies with sparse spatial extent.

The state-of-the-art for automated brain aneurysm detection in TOF-MRA has been rapidly advancing in the last five years, especially due to the advent of deep learning algorithms. However, further multi-site validation is needed before safely applying these algorithms during routine clinical practice. Although (Joo et al., 2020; Ueda et al., 2019) did publish results obtained from multiple institutions, none of them released their dataset publicly which makes comparisons between algorithms unfeasible. The comparisons between methods are further hindered by the use of non-standardized evaluation metrics (e.g. FROC/lesion-wise sensitivity/subject-wise specificity) or by the fact that not all related studies include both patients (subjects with aneurysms) and controls (subjects without aneurysms). By openly releasing our dataset, we aim to bridge the data availability gap and foster reproducibility in the medical imaging community. The combination of our in-house dataset and the ADAM dataset will allow researchers to assess the realistic robustness of proposed algorithms on heterogeneous data generated from different scanners, acquisition protocols and study population. In addition, it could help increasing detection performances which are still too far from being clinically useful, considering that even the team with highest sensitivity on the ADAM test set (team *xlim*) only reaches a value of 70% (i.e., 30% of aneurysms still not detected), with 4 FPs per case.

In a separate analysis, we also computed the sensitivity of our model with respect to the PHASES score risk of rupture, location, and size. No significant differences were found across the three groups indicating that our model is robust to different aneurysm types. However, when stratifying the aneurysm sizes into finer subgroups, we noticed that sensitivity for extremely tiny aneurysms ($\leq 3$ mm) was significantly lower, which confirms a known trend (Timmins et al., 2021).

Our work has several limitations. First, even combining our in-house dataset with the ADAM dataset, the number of subjects is still limited when compared to some related TOF-MRA (Joo et al., 2020; Ueda et al., 2019) or Computed Tomography Angiography (Park et al., 2019; Shi et al., 2020; Yang et al., 2020) studies. Second, we acknowledge that the number of patients for whom we compared the different annotations schemes (i.e., weak vs. voxel-wise) is limited (N = 38); it is possible that statistically significant performance differences could be found with a larger sample size. Third, we have to further increase detection performances if we plan to deploy our model as a second reader for radiologists, especially to detect tiny aneurysms which are more frequently overlooked (Keedy, 2006).

In future works, we aim at enlarging the TOF-MRA dataset and experiment new variants of the 3D encoding–decoding UNET. For instance, we might consider a multi-scale approach with patches of larger (or smaller) scales. Alternatively, we are considering combining our anatomically-driven approach with the novel nnUnet model (Isensee et al., 2021) which has proven to be effective not only for aneurysm detection (it was adopted by 2 of the top-performing teams in the ADAM challenge), but also for several other segmentation tasks. We believe this combination holds potential to boost detection performances. Also, the ablation study performed in the Online Resources – Section F showed that pre-training on the ADAM dataset did not increase detections results. Therefore, future works should investigate a different transfer learning approach to better leverage knowledge acquired from the ADAM dataset. Last, we plan to conduct further error analyses to identify common patterns for both false positive and false negative cases.

In conclusion, our study presented an anatomically-informed 3D UNET that tackles brain aneurysm detection across different sites. The combination of time-saving weak labels and anatomical prior knowledge allowed us to build a robust deep learning model. We believe our approach and dataset (both openly available) can foster the development of clinically applicable automated systems for the task at hand.

## Information Sharing Statement - Data Availability

Our open-access dataset is available on OpenNeuro under the CC0 license at https://openneuro.org/datasets/ds003949. The ADAM dataset can be downloaded from the challenge website https://adam.isi.uu.nl/data/ after signing a confidentiality agreement. The code used for this study is available at https://github.com/connectomicslab/Aneurysm_Detection under the Apache-2.0 license, together with the configuration files to replicate all the experiments, and the weights of the trained model if users simply want to perform inference.

**Supplementary Information** The online version contains supplementary material available at https://doi.org/10.1007/s12021-022-09597-0.

**Acknowledgements** We would like to thank the organizing team of the ADAM challenge for their great effort and availability.

**Funding** Open access funding provided by University of Lausanne.

**Publisher's Note** Springer Nature remains neutral with regard to jurisdictional claims in published maps and institutional affiliations.